# Same data may bring conflict results: a caution to use the disruptive index

Guoqiang Liang, Yi Jiang, Haiyan Hou

*WISE Lab, Dalian University of Technology, Dalian, Liaoning 116024, China*

**Abstract:**

In the last two decades, scholars have designed various types of bibliographic-related indicators to identify breakthrough-class academic achievements (Lutz Bornmann, Tekles, Zhang, & Ye, 2019). In this study, we take a further step to look at properties of the promising "disruptive index" (DI), proposed by Wu et al. (Wu, Wang, & Evans, 2019), thus deepening our understanding of DI and further facilitating its wise use in bibliometrics. Using publication records for Nobel laureates between 1900 and 2016 (Li., Yin., Fortunato., & Wang., 2019), we calculate the DI of Nobel Prize-winning articles and its benchmark articles in each year and use the median DI to denote the central tendency in each year, and compare results between Medicine, Chemistry, and Physics. We find that conclusions based on DI depend on the length of their citation time window, and different citation time windows may cause different, even controversial, results. Also, discipline and time play a role on the length of citation window when using DI to measure the innovativeness of a scientific work. Finally, not all articles with DI equals to 1 were the breakthrough-class achievements. In other words, the DI stands up theoretically, but we should not neglect that the DI was only shaped by the number of citing articles ($n_i+n_j$) and times the references have been cited ($n_k$), these data may vary from database to database.

**One Sentence Summary:** After analyzing the disruptive index (DI) of Nobel laureates' publishing records, we find three pitfalls of using this indicator: the DI depends at least on the length of the citation window, discipline and publishing year of the work, and not all papers with DI equals to 1 have the highest disruptiveness effect.

**Main Text:**

*Background:* Foundational theories of innovation distinguish between two types of innovation. The first works within a well-defined tradition, where innovation builds upon and enhances its predecessors and, therefore, results in consolidation (Foster, Rzhetsky, & Evans, 2015; Funk & Owen-Smith, 2017). The second type challenges the existing paradigm and has the potential to disrupt existing fields (Kuhn, 1977; Wu et al., 2019). Despite the substantive and theoretical importance of differentiating between these two types of innovations, few studies measure them quantitatively; however, in the last two decades, scholars have designed various types of bibliographic-related indicators to identify breakthrough-class academic achievements (Lutz Bornmann et al., 2019), such as Uzzi et al. (Uzzi, Mukherjee, Stringer, & Jones, 2013), who applied the Z-score to measure the balance between atypical and conventional combination, which is a proxy of innovation. Following this, Foster et al. (Foster et al., 2015) analyzed millions of biomedical abstracts and developed a typology network of scientists' research strategies. Wang et al. (Wang, Veugelers, & Stephan, 2017) introduced a "novelty score" to measure the innovativeness of articles based on cosine similarity. Wu et al. (Wu et al., 2019) designed a "disruptive index" (DI) to measure the disruptiveness of a scientific work.

According to Wu et al. (Wu et al., 2019), DI was used in evaluating works introducing something new to their field to capture the degree to which this work disrupts its field and eclipses the work upon which it had been built.

For example, Nobel Prize-winning articles (NPs) illustrate that DI can better distinguish NPs from other articles. A recent study (L. Bornmann & Tekles, 2019) that analyzed four papers and has shown that DI depends on the citation window, which must include at least three years to produce meaningful results. In this study, we take a further step to look at properties of the promising DI, proposed by Wu et al. (Wu et al., 2019), thus deepening our understanding of DI and further facilitating its wise use in bibliometrics.

*Data collection and methodology:* We use the publication records for Nobel laureates between 1900 and 2016 (Li. et al., 2019) as the NPs in this study. All cited-citing relations of NPs were derived from Web of Science (WoS). The benchmark articles (BPs) are then implemented by randomly selecting another paper in the same journal issue for each of the NPs. Finally, 646 NPs and 653 BPs were collected after deleting the papers missing reference information. We calculate the DI of NPs and BPs in each year and use the median DI to denote the central tendency in each year, as the data do not follow a normal distribution. Due to the word limit, we do not describe the calculation of DI; however, the reader can find these details in our references (Wu et al., 2019).

*Results - DI depends on citation time window:* Fig. 1 shows that DI presents an increasing tendency, in general, but different results occur when the citation time windows (t) were set among: t≤5, 6≤, t≤9, and t≥10. (As our data do not follow a normal distribution, mean DIs do not convey much information.)

*Results - Disciplinary and time effect on DI:* Fig. 2A presents that the Physics, Medicine, and Chemistry field have different DI citation window variation tendency of NPs and BPs. The DI of NPs in Medicine is less than BPs published within 10 years or longer. The DI of NPs in Chemistry is less than BPs published within 6 years or longer, and the DI of NPs in Physics is less than BPs published within 4 years or longer. As for the time effect on DI, after considering the coverage of WoS, we compared the papers published before 1980 versus those published after 1980. Fig. 2 illustrates the DI dependence of time is different for NPs and BPs published before 1980 and after 1980.

*Conclusions:* **Firstly, conclusions based on DI depend on the length of their citation time window, and different citation time windows may cause different, even controversial, results.** Disruptive achievements usually require more time to be recognized; therefore, a 10-year citation window, at least, will be more suitable than our reference (L. Bornmann & Tekles, 2019) proposed. **Secondly, disciplinary and time effects may exist when using DI to measure the innovativeness of a scientific work.** The physics, medicine, and chemistry field have different DI variation tendency for the NPs and BPs. **Finally, not all articles with DI equals to 1 were the breakthrough-class achievements,** because DI only shaped by the number of citing articles ($n_i+n_j$), number of references, and the number of times the references have been cited ($n_k$). We found many BPs have DI that equals to 1 in each year after they published (such as paper ID=WOS:000200263700094).

**Acknowledgement:**
The authors would like to thank Yi Bu for helpful discussions.

**References and Notes:**
Bornmann, L., & Tekles, A. (2019). Disruption index depends on length of citation window. *El Profesional de la Información,, 28*(2), e280207.
Bornmann, L., Tekles, A., Zhang, H. H., & Ye, F. Y. (2019). Do we measure novelty when we analyze unusual combinations of cited references? A validation study of bibliometric novelty indicators based on F1000Prime data. *Journal of Informetrics, 13*(4). doi:10.1016/j.joi.2019.100979

Foster, J. G., Rzhetsky, A., & Evans, J. A. (2015). Tradition and Innovation in Scientists' Research Strategies. *American Sociological Review, 80*(5), 875-908. doi:10.1177/0003122415601618

Funk, R. J., & Owen-Smith, J. (2017). A Dynamic Network Measure of Technological Change. *Management Science, 63*(3), 791-817. doi:10.1287/mnsc.2015.2366

Kuhn, T. S. (1977). *The Essential Tension: Selected Studies in Scienific Tradition and Change*. Chicago: University of Chicago Press.

Li., J., Yin., Y., Fortunato., S., & Wang., D. (2019). A dataset of publication records for Nobel laureates. *Sci Data, 6*(1), 1-10. doi:10.1038/s41597-019-0033-6

Uzzi, B., Mukherjee, S., Stringer, M., & Jones, B. (2013). Atypical combinations and scientific impact. *Science, 342*(6157), 468-472. doi:10.1126/science.1240474

Wang, J., Veugelers, R., & Stephan, P. (2017). Bias against novelty in science: A cautionary tale for users of bibliometric indicators. *Research Policy, 46*(8), 1416-1436. doi:10.1016/j.respol.2017.06.006

Wu, L., Wang, D., & Evans, J. A. (2019). Large teams develop and small teams disrupt science and technology. *Nature, 566*(7744), 378-382. doi:10.1038/s41586-019-0941-9

**Figures:**

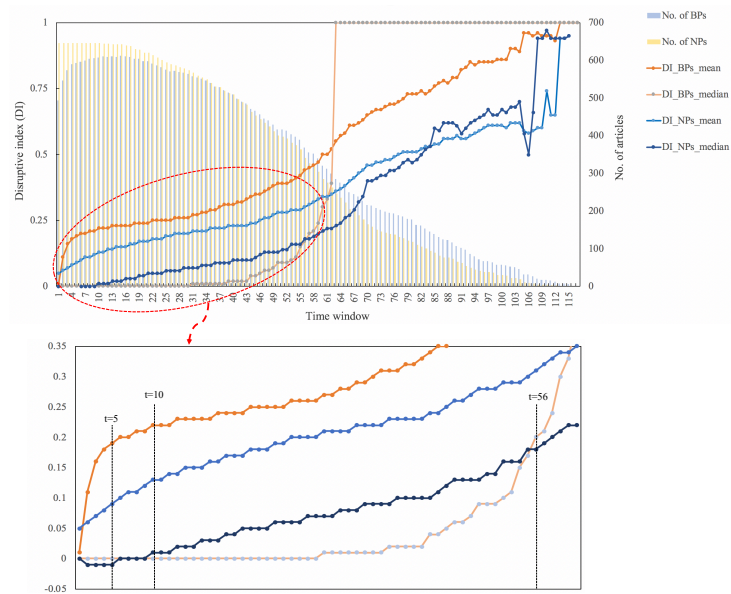

**Fig. 1** DI of NPs and BPs after their publication in each year

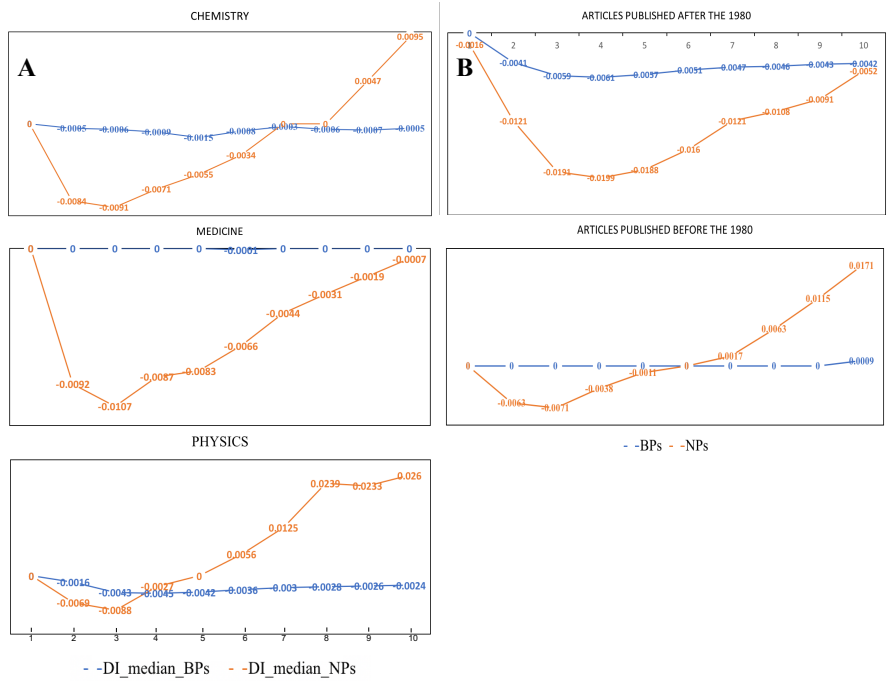

**Fig. 2** Disciplinary and time different on DI's length of citation window